\documentclass[aip]{revtex4-1}

\usepackage{graphicx}
\usepackage{dcolumn}
\usepackage{bm}

\usepackage[utf8]{inputenc}
\usepackage[T1]{fontenc}
\usepackage{mathptmx}
\usepackage{etoolbox}

\usepackage{subcaption}
\usepackage{todonotes}

\makeatletter
\def\@email#1#2{%
 \endgroup
 \patchcmd{\titleblock@produce}
  {\frontmatter@RRAPformat}
  {\frontmatter@RRAPformat{\produce@RRAP{*#1\href{mailto:#2}{#2}}}\frontmatter@RRAPformat}
  {}{}
}%
\makeatother
\begin{document}


\title[]{Parametric excitations of coupled nanomagnets}
\author{Domonkos Laszlo Farkas }
\author{Gyorgy Csaba}
 \email{csaba.gyorgy@itk.ppke.hu}
\affiliation{ Faculty of Information Technology and Bionics, Pazmany University Budapest, Hungary }%

\date{\today}

\begin{abstract}
We demonstrate that parametrically excited eigenmodes in nearby nanomagnets can be coupled to each other. Both positive (in-phase) and negative (anti-phase) couplings can be realized by a combination of appropriately chosen geometry and excitation field frequency. The oscillations are sufficiently stable against thermal fluctuations. The phase relation between field-coupled nanomagnets shows a hysteretic behavior with the phase relation being locked over a wide frequency range. We envision that this computational study lays the groundwork to use field-coupled nanomagnets as parametrons as building blocks of logic devices, neuromorphic systems of Ising machines.
\end{abstract}

\maketitle

\section{Introduction}

Our research is motivated by the possibility of using the oscillation phase of a nanomagnet as the information-carrying physical variable in a computing system. The idea of using the phase of a parametrically-excited system as a variable in computing has a long history \cite{ref:wigington} -- in fact the magnetic parametron was one of the first practically successful computing paradigms before the age of integrated, transistor-based digital electronics. 

In this paper - following the footsteps of earlier works \cite{ref:parametron,ref:bauer} -- we first show that small-sized nanomagnets can indeed act as parametrons. It means that they produce $f_0$ frequency oscillations  to a $2f_0$ frequency excitation and  be locked to this external pumping field with two distinct, stable phases. 

To use individual magnetic parametrons as building blocks of computing devices they should  be interconnected with each other. One possibility is to use the stray field of the magnets for this, which comes for free, not requiring additional hardware. The oscillatory stray field of a neighboring magnet shifts the resonance frequency of a node in such a way that a given $2f_0$ excitation may favorably excite only on or the other phase, effectively realizing positive (in-phase) or negative (anti-phase) couplings between the phase variables of the magnets.

Nanomagnets coupled by their  stray fields  have been intensely researched for  applications in beyond-Moore computing paradigms. For example, in Nanomagnet Logic (NML) the magnetic orientation of a nanomagnet acts as a logic variable \cite{ref:nml}, and this magnetic orientation, in turn, controlled by the field-coupling to neighboring magnets. The NML architecture enables a potentially low-power, non-volatile, straightforwardly realizable architecture for Boolean and non-Boolean functions. In this paper we extend the NML paradigm to dynamically-coupled nanomagnets. We call the field-coupled building blocks Coupled NanoMagnet Parametrons (CNMPs).


The realization of computationally useful magnetization patterns requires both negatively (anti-phase) and positively (in-phase) coupled magnets or at least negatively coupled ones. In-phase coupled magnets alone are not computationally useful as they have only a trivial ground state with all magnets oscillating in phase. We will show that the strength and the sign of couplings can be engineered by choosing the right parametric excitation frequency  for a given geometry.

\section{Oscillatory states of a stand-alone nanomagnet}

Oscillatory eigenstates of nanoscale  magnets are a well-studied and the reader is referred to \cite{ref:serpico} for more information. Here we are primarily interested in parametrically excited quasi-uniform oscillation modes in small magnets as they are the ones that provide sufficiently strong fields for coupling to neighbors.

\subsection{Simulation setup}   

\begin{figure}[htpb!]
    \begin{center}
        \includegraphics[width=\columnwidth]{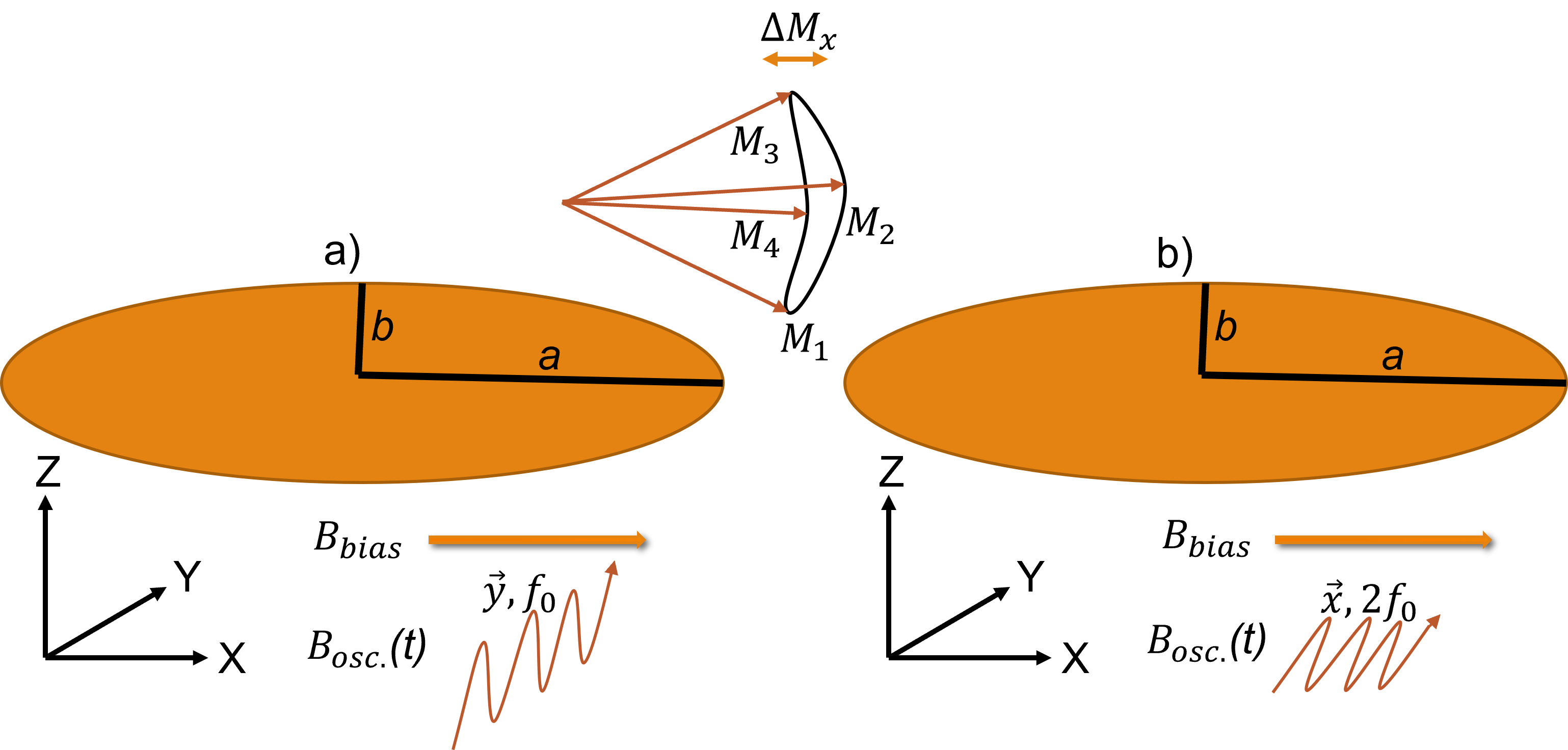}
        \caption{Geometry for exciting the mode with eigenfrequency $f_0$ of an ellipse-shaped magnet (2b=a). In $a)$, the eigenmode is excited by its eigenfrequency with a field perpendicular to the bias field (and thus falls in the same plane as the oscillation). $b)$ As the spins' precession orbit is elliptical in this eigenmode, it is also possible to parametrically pump it with a double-frequency ($f=2f_0$) excitation along the bias field's direction.}
        \label{fig:setup}
    \end{center}
\end{figure}

 We study an ellipse-shaped (major axis=160 nm, minor axis=80 nm, thickness=20 nm) YIG nanomagnet,  see Figure \ref{fig:setup}. The elliptical shape is useful to increase the ellipticity of the magnetization precession orbit and remove the degeneracy of eigenmodes. A static bias field of 500 mT is applied along the easy axis of the nanomagnet. Additionally, a weaker sinusoidal field is applied to generate oscillations of the magnet. The sinusoidal field is aligned  parallel to the bias field for parametric excitation and perpendicular to the bias field in case of direct excitation of a mode.

The relatively small nanomagnet size is chosen for two reasons. The quasi-uniform oscillation creates the strongest stray field so the strongest couplings are expected to nearby magnets. In addition to this, for such small nanomagnets only the lowest-energy eigenmode is accessible with a straightforwardly applicable $f<20$ GHz excitation frequency - this simplifies the physical picture by removing the internal degrees of freedom. 

For the calculations the Mumax simulation code \cite{ref:mumax} is used with standard YIG parameters ($M_{s}$ = 1.40E+5 A/m, $A$ = 3.65E-12 J/m, $\alpha$ = 0.0005) and we assume that this thin film is precisely modeled by a two-dimensional simulation (single layer of computing cells along $z$ axis, but with 20 nm height). We used the temperature-dependent module of Mumax and most simulations (unless otherwise indicated) done at room temperature ($T=300$ K). It simulates thermal fluctuations by adding a white-noise-like effective magnetic field to all the computing cells. The higher the temperature, the more prominent the fluctuations becomes. Most simulations were repeated multiple times using different seeds for random thermal field generation - this removes simulation artifacts and ensures the effects we see are robust against thermal noise. Thermal effects do influence the results: for example, parametric excitations require significantly higher excitation power than they would require at zero temperature.

\subsection{Localizing eigenfrequencies of modes}

Eigenfrequencies of a nanomagnet are determined by the effective magnetic field, which  includes the magnet's own demagnetizing field and external field sources. 

We use a high-bandwidth pulse (impulse response) and the thermal spectrum to identify eigenmode frequencies. Both methods  excite a wide range of frequencies, thus they excite all eigenfrequencies at once. Applying Fourier transform (FFT) on the time-domain magnetization dynamics will show them as peaks in the spectral domain. Although the two exciting methods may provide similar results, looking at thermal fluctuations have some benefits. First, it has more uniform spectral density so it excites all eigenfrequencies more evenly. Secondly, it can be prolonged at any scale (e.g. for better FFT resolution) with the same effectiveness, whereas the impulse response results in an exponential decay. Thirdly, the high magnetic field of the impulse may distort the system's behaviour, as eigenfrequencies depend on the magnetic field. 


Eigenmodes may alternatively may identified by sweeping the frequency over time and determining the frequency where maximum-amplitude oscillations occur - this method is useful if the scanned frequency range is narrow.

\subsection{Excitation of eigenmodes, parametric pumping}

Parametric excitation of nanomagnet eigenmodes is widely studied, see e.g. \cite{ref:demidov, ref:efield}. Here we focus on perhaps the simplest case of a parametric process, the excitation of a quasi-uniform mode, leaving the oscillation phase a free variable. 


A simple picture of parametric excitation is illustrated in Figure \ref{fig:setup}b contrasting it to the direct excitation of \ref{fig:setup}a. In direct excitation, a resonant excitation field is applied perpendicular to static magnetization, maximizing the torque. Although resonant excitation requires lower excitation power, the resulting oscillation does not have a degree of freedom such as two possible phases for parametric pumping. 

\begin{figure}[htpb!]
\centering
\begin{subfigure}{0.5\columnwidth}
  \centering
  \includegraphics[scale=0.25]{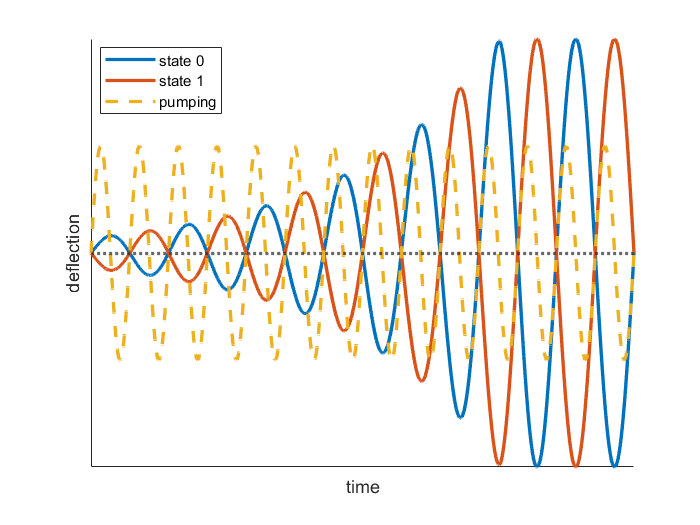}
  \caption{Artificial buildup of a parametron.}
  \label{fig:artificial_parametric}
\end{subfigure}%
\begin{subfigure}{0.5\columnwidth}
  \centering
  \includegraphics[scale=0.25]{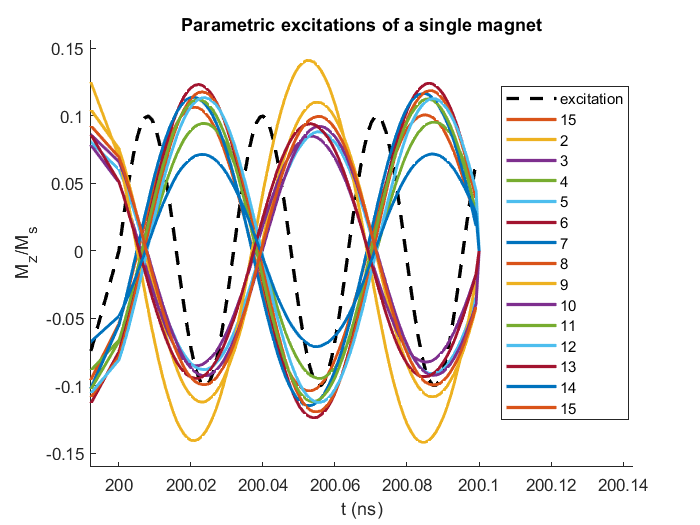}
  \caption{Resulting CNMP oscillations using different random seeds.}
  \label{fig:parametric_simulated}
\end{subfigure}
\caption{Parametric pumping allows oscillations with two different phase, and the same amplitude. Subfigure b) shows that oscillation phase is more robust to thermal fluctuations than amplitude. The magnet's major axis was 160 nm, the minor 80 nm and the magnetization precesses in a coherent way.}
\label{fig:pumping_setup}
\end{figure}

In parametric excitation, the ellipsoidal-shaped nanomagnet is excited parallel to the bias field with an oscillatory magnetic field of $B_{osc}$ (typically in the 25 mT range) and frequency $f=2f_0$. As Figure \ref{fig:setup} shows, a precessional motion with $f_0$ frequency and non-zero ellipticity will have a component oscillating at $2f_0$ frequency along the biased easy axis. The parametric pumping will couple to this component and sustain the precession against dispersion losses. As illustrated in Figure \ref{fig:pumping_setup}a), a small  eigenoscillation (buried in thermal noise) occurring at $f_0$ will be amplified to large amplitudes.  Figure \ref{fig:pumping_setup}a) is an illustration, and Figure \ref{fig:pumping_setup}b) is a result of actual simulations of an ellipsoidal nanomagnet, starting from different, temperature-induced random states.  It typically takes several hundred oscillation cycles to reach the steady-state oscillation amplitude.  

\begin{figure}[htpb!]
    \begin{center}
        \includegraphics[scale=0.25]{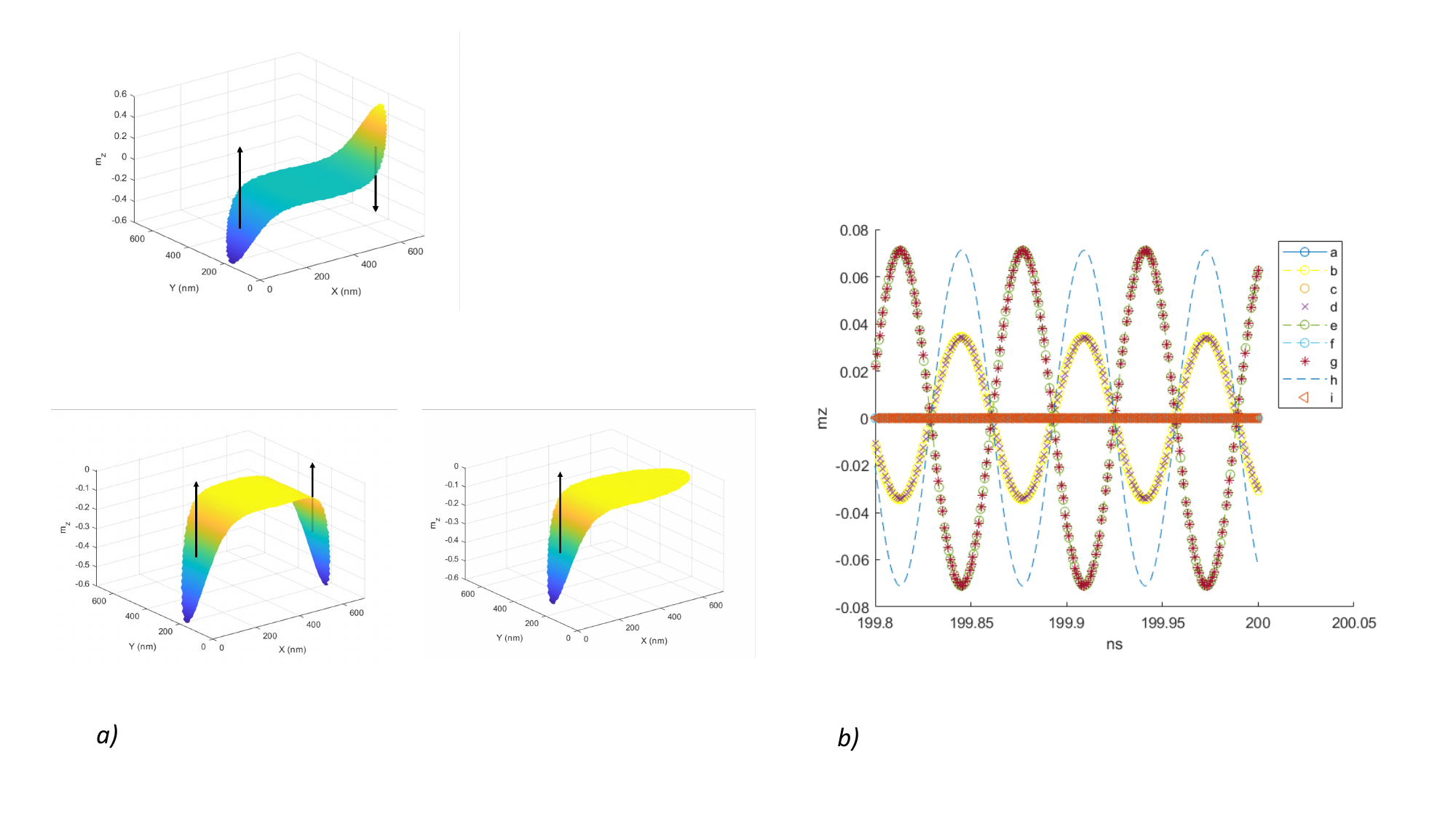}
        \caption{a) In case of larger-sized magnets, modes can be localized on one side of the magnet. Thus parametrically excited eigenmodes may form with different amplitude due to this internal degrees of freedom. The size of the magnet was 640 nm x 320 nm, temperature was set to zero for these simulations.}
        \label{fig:pumping_big}
    \end{center}
\end{figure}

A key feature of parametric excitation, as shown in Figure \ref{fig:pumping_setup}, is that the oscillations of the magnet  may occur in one of exactly two possible locked phases with respect to the excitation frequency. In stand-alone magnets the phase is 'decided' randomly by the initial random state. It is worthwhile to note that in-larger sized magnets, non-uniform modes will be excited by parametric pumping, and they add additional degrees of freedom to the system. So locking to the subharmonic of the excitation could happen in multiple ways as shown in Figure\ref{fig:pumping_big}. We chose the size of the nanomagnet in the deep submicron regime to suppress the formation of these modes.

\section{Mode coupling in coupled nanomagnets}

In case of two magnets placed next to each other, the stray field shifts eigenfrequencies. We can obtain the modified system's eigenfrequencies with the same methodology as for a stand-alone magnet. First we used the thermal fluctuations as a uniform, high-bandwidth excitation, which excited all modes at once. Then we applied FFT on the sampled average magnetization of each magnet separately to localize the eigenfrequencies. 

We observed that the presence of another magnet clearly affects the eigenfrequencies. In case the magnets were placed next to each other along their hard axis, the eigenfrequencies were lowered by a few tens of MHz. On the other hand, when the magnets' easy axis coincided, the effect was the opposite, the eigenfrequencies increased. In both cases, the frequency shift decreases as the pair's distance increases, converging  to the stand alone magnet case. Additionally, we see new peaks, which faded as the gap increased. 

\begin{figure}[htpb!]
    \begin{center}
    \includegraphics[scale=0.25]{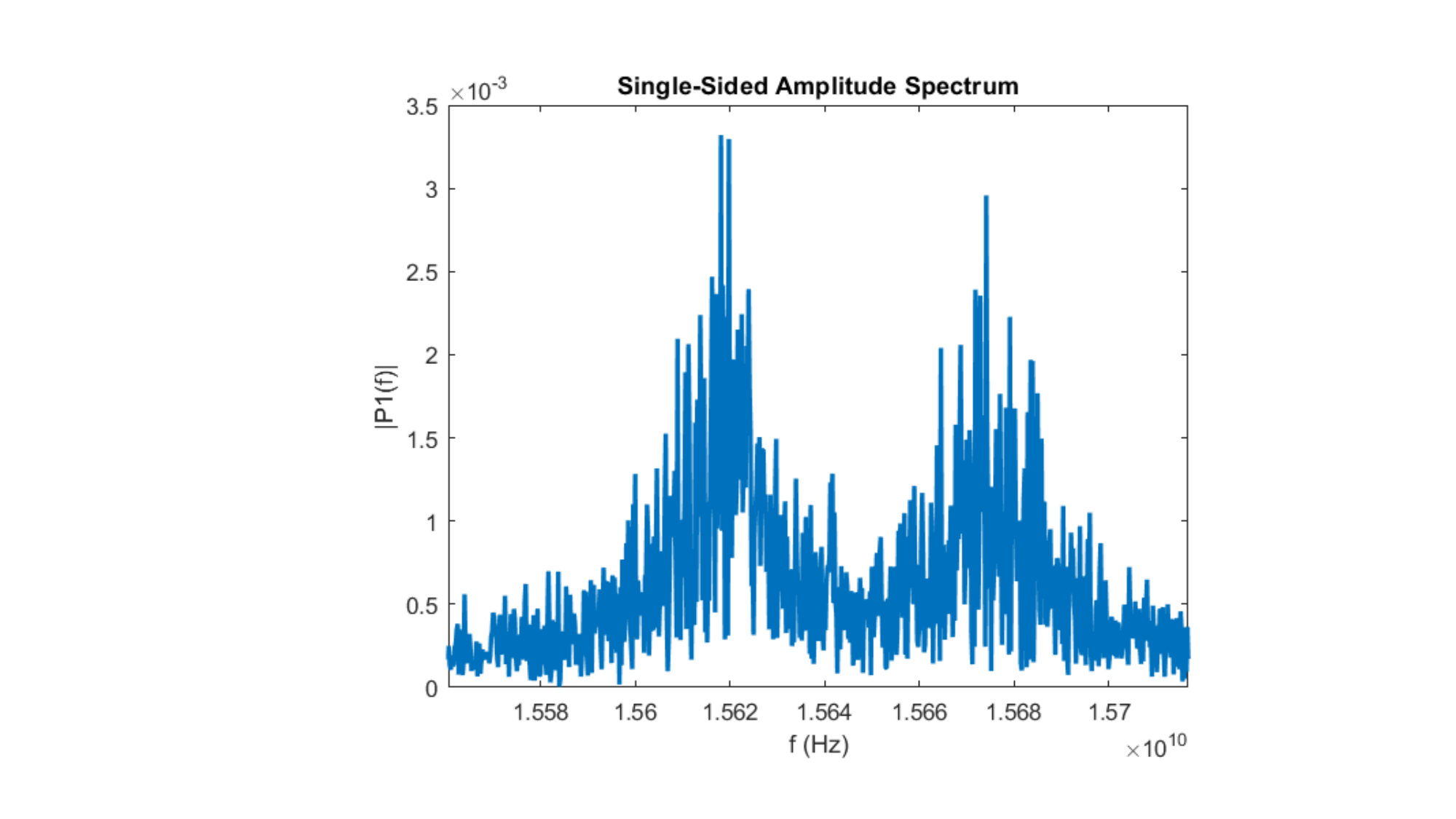}
    \caption{The geometry of two couples nanomagnet and their thermal spectrum. The peaks correspond to eigenmodes coupled in different phases. The magnets were placed next to each other along their hard ($y$) axis.}
    \label{fig:thermal_spectrum}
    \end{center}
\end{figure}

 Figure \ref{fig:thermal_spectrum} shows the obtained frequency-domain around the first eigenfrequency in a constellation, where the magnet pair was placed next to each other along their hard axis (other eigenfrequencies' or the other constellation's domain is quite similar, only higher in frequency). Two separate peaks, appear of which the second one diminishes if we apply FFT on the full system's average magnetization rather than separately on each magnet. This indicates that the second peak corresponds to the anti-phase oscillation of the magnets, where they cancel out on average, whereas the first peak corresponds to the same-phase oscillation. 

If only thermal excitations drive the system, then the the two phases are present simultaneously. If parametric pumping is applied, then one or the other phase configuration (which is closer to $f_0$ will be amplified.

It is worthwhile to note that the modes' spatial distribution slightly changes compared to the stand alone magnet setup, they become more non-uniform, there is a gradual increase in amplitude going from the magnets' close ends towards their far ends.

\subsection{Sweeping the frequency around the resonance}

Coupling of the magnets was studied by extracting their phases from the mumax simulations. First we calculated the volume-averaged magnetization for the two neighboring magnets. The magnetization values were normalized in the sliding window, by the ratio of the maximum value (per magnet) and a predefined constant. This step removes the amplitude-changes of the oscillations. Secondly, we took the dot product of the two magnet's normalized values in the window. This dot product is positive for in-phase oscillations, as all members are a multiplication of two \textit{same}-sign values, and it is negative for anti-phase oscillations, as all members are a multiplication of two \textit{different}-sign values. There is a strictly monotonic transition between the two.

\begin{figure}[htpb!]
\centering
\begin{subfigure}{0.5\columnwidth}
  \centering
  \includegraphics[scale=0.25]{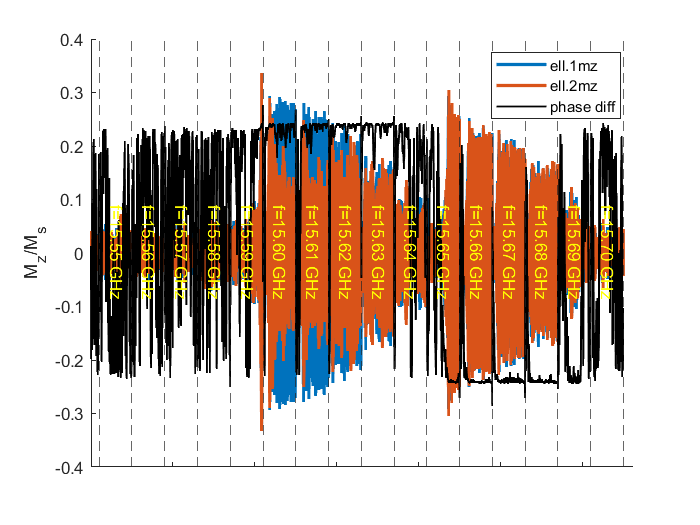}
  \caption{}
  \label{fig:short20nm}
\end{subfigure}%
\begin{subfigure}{0.5\columnwidth}
  \centering
  \includegraphics[scale=0.25]{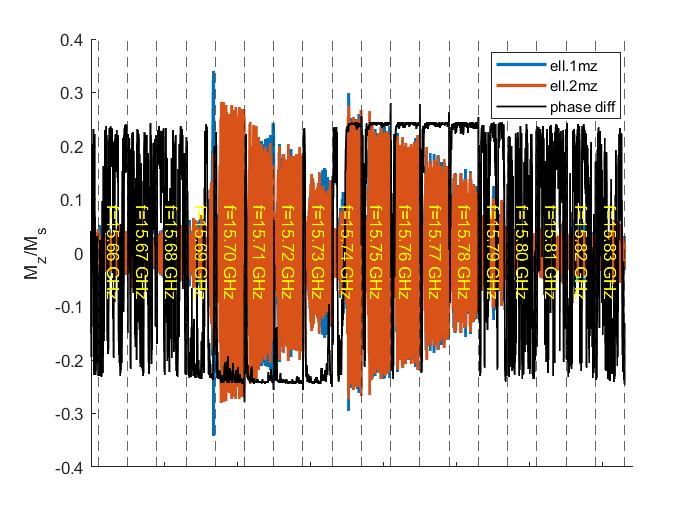}
  \caption{}
  \label{fig:long20nm}
\end{subfigure}
\caption{Phase correlation of magnets at different frequencies for hard (a) and easy (b) axis constellations. the black line is an indicative phase-relation of the two magnets oscillation while the yellow texts in the middle show the excitation frequencies at different time intervals. The magnetization is relaxed (without RF driving) to a thermal state after each magnetization interval.}
\label{fig:coupling_domains_two_constellations}
\end{figure}

\begin{figure}[htpb!]
\centering
\begin{subfigure}{0.5\columnwidth}
  \centering
  \includegraphics[scale=0.25]{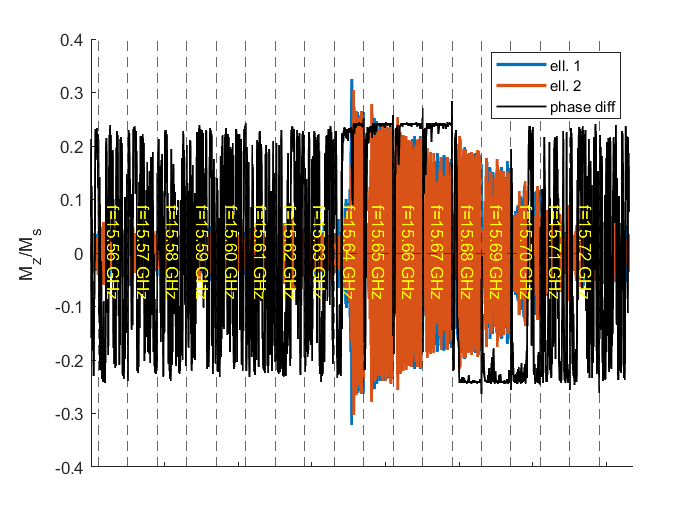}
  \caption{}
  \label{fig:short80nm}
\end{subfigure}%
\begin{subfigure}{0.5\columnwidth}
  \centering
  \includegraphics[scale=0.25]{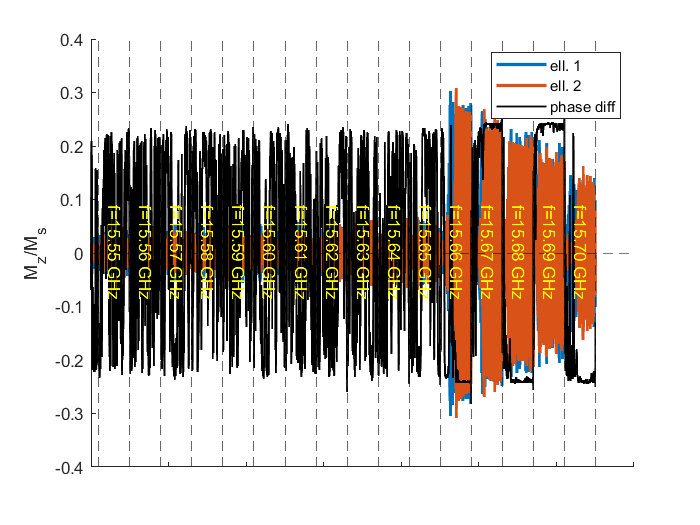}
  \caption{}
  \label{fig:short1000nm}
\end{subfigure}
\caption{Phase correlation of magnets placed in different distances. a) For far-away magnets (80 nm), there is a relatively narrow  frequency range where the pumped magnets are in a definite phase-relation. This frequency regime opens to much wider range when the magnets get closer to each other. b) For even further-away magnets (1000 nm), the phase-relation is random and the eigenfrequency shifts back to the stand alone magnet case, i.e. the magnets don't affect each other.}
\label{fig:sweep_distance}
\end{figure}

Exciting the CNMP with different excitation frequencies around the stand-alone magnet eigenfrequency showed that depending on the frequency, both negative (anti-phase)  and positive (in-phase) couplings may occur. A clear boundary separates the positive and negative couplings' regimes. Figure \ref{fig:coupling_domains_two_constellations} shows this separation, using the above-mentioned indicative phase difference evaluation. The figure also illustrates that the easy and hard axis constellations' coupling regimes look similar, but the positive-negative coupling regimes are reversed, i.e. the order in frequency is the following: hard axis constellation's positive coupling regime (15.60 GHz - 15.64 GHz), hard axis constellation's negative coupling regime (15.65 GHz - 15.69 GHz), stand-alone magnet's eigenfrequency peak (15.67 GHz - 15.69 GHz), easy axis constellation's negative coupling regime (15.69 GHz - 15.73 GHz), easy axis constellation's positive coupling regime (15.74 GHz - 15.79 GHz). So in both cases the negative couplings are closer to the stand-alone eigenfrequency than the positive couplings. 

This separation can be explained by the eigenfrequencies' dependence on the magnetic field. The two magnet's field affect each other's effective field, therefore its eigenfrequency as well. However this change depends on the coupling phase, because the magnets demagnetizing field's components and the other magnet's effect may sum up or cancel out depending on the sign of the coupling. So that the positive and negative couplings have different eigenfrequencies. The negative couplings might be closer to the stand-alone eigenfrequency because more components cancel out, or the summed up components have opposite effects on the eigenfrequencies.

The gap's size between the magnets are crucial in regard of the coupling regime's separation. Figure \ref{fig:sweep_distance} shows the same kind of simulation with a wider gap (80 nm) and with a much wider gap (1000 nm) compared to the previous results on Figure \ref{fig:coupling_domains_two_constellations} (20 nm). One can see that the two regimes are more pushed together at 80 nm, and completely coincide at 1000 nm (there is no separation). Furthermore the resonant domain gets closer to the stand-alone eigenfrequency. These observations are not surprising: as the gap size increases the magnet's effect on each other's effective field decreases, so they are getting closer to the stand-alone magnet case. 

If the magnets are too far away then  there is no coupling between them and they randomly 'choose' one of the two possible phases (forced by the parametric excitation). Their phase difference is only stable because the individual oscillations (parametrons) cannot switch phase after they reached a sufficiently large amplitude. But in different simulations that use different seeds for thermal field generation  the positive-negative "coupling" pattern would be completely different.

There might be a switch in case of coupling around the separation border, where the lower-frequency coupling has small steady-state oscillation amplitude. Thermal fluctuations can switch this small-amplitude oscillation into the other coupling, but for the reverse switch, much higher fluctuations are required, as the oscillation amplitude is much higher, so that more stable to noise.
    
Finally a general rule is that, for any constellation and coupling, lower frequencies - if still in the excitable regime- result in higher amplitude oscillations. This is also due to the eigenfrequency - magnetic field dependence and will be discussed in more detail in the following section.

\section{Hysteretic behavior of phase coupling}

In the previous sections (Figures \ref{fig:coupling_domains_two_constellations}) the frequency was changed by allowing the magnets to relax (i.e. their phase to randomize) between the steps of the parametric excitations. 

\begin{figure}[htpb!]
\centering
    \begin{subfigure}{0.5\columnwidth}
        \centering
        \includegraphics[scale=0.25]{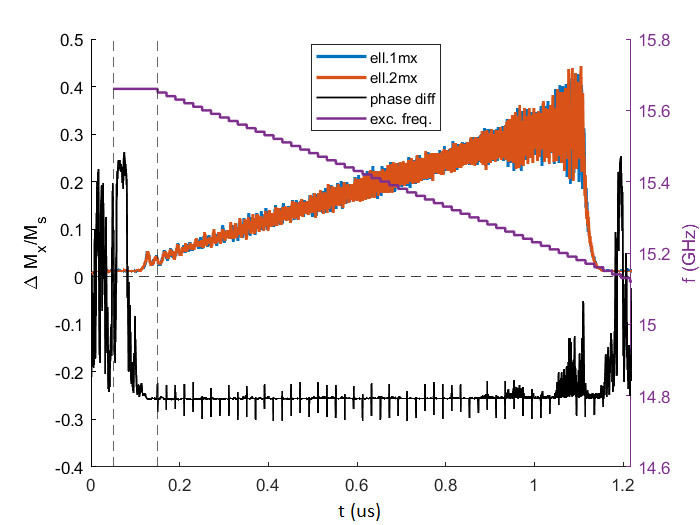}
        \caption{}
        \label{}
    \end{subfigure}%
    \begin{subfigure}{0.5\columnwidth}
      \centering
      \includegraphics[scale=0.25]{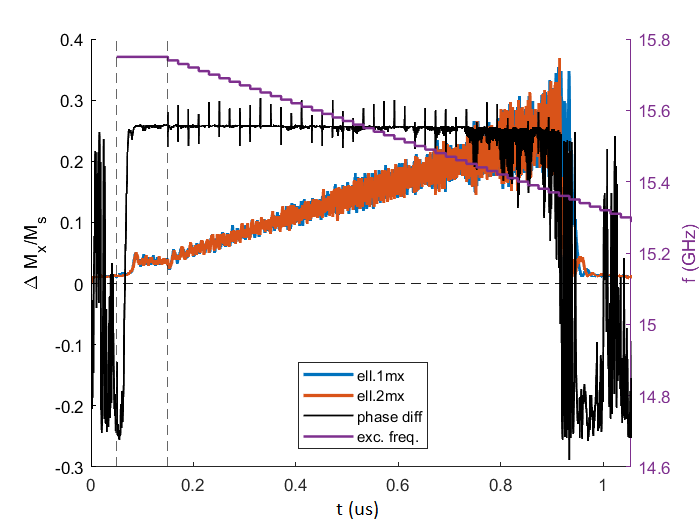}
      \caption{}
      \label{}
    \end{subfigure}
    \caption{Oscillation amplitude of magnet pairs while sweeping the excitation frequency (purple stepwise line) downwards. The initial frequency was set such that it's in the middle of the higher-frequency coupling regime, i.e. negative coupling for the hard axis constellation (a), positive coupling for the easy axis constellation (b). In both cases, the initial coupling was preserved through a couple of tenth gigahertz wide frequency domain which is a multiple of the width of the excitable domain without hysteresis. Note that the evenly-spaced spikes in the phase difference are only visualization artifacts.}
    \label{fig:coupling_hysteresis}
\end{figure}

Here we study the oscillation's behaviour during a frequency sweep without relaxing the magnets at frequency steps. The results are shown in Figure \ref{fig:coupling_hysteresis}. The phase relation shows a hysteretic behavior: a phase relation, once formed between the magnets, remains stable  over a wide range of frequencies, even at frequencies where the opposite phase would be energetically favorable.

Apparently, CNMPs have a similar hysteresis than what is found in large-amplitude oscillations of stand alone magnets \cite{ref:loubens}: stabilized oscillations can stay in the same coupling while sweeping through the other coupling's regime and in much lower frequencies, where the pair cannot be excited in either phase without hysteresis. Even more the oscillation amplitude increases quite notably until a sharp decay. 

Even a parametrically excited stand-alone magnet can hold its phase once the oscillation is strong enough, so it is reasonable that CNMPs do not switch either, especially to a lower amplitude excitation.


The oscillation can retain and even increase its amplitude because of the magnetic field dependence of the eigenfrequencies, notably the magnets' demagnetizing fields. The greater the oscillation, the smaller the magnetization component along the easy axis which lowers the eigenfrequency. Thus, if the frequency is decreased slowly, the steady-state oscillation amplitude increases and shifts the eigenfrequency downwards, so it stays close to the excitation's frequency. At high amplitudes it starts to become less stable (see the increasing amplitude deviations in Figure \ref{fig:coupling_hysteresis} at high amplitudes), and eventually the excitation cannot counter the growing dispersion losses and fast decay starts. The decreasing oscillation shifts back the eigenfrequency, further from the excitation's frequency, so the decay even accelerates itself.

Note that this hysteresis behaviour is much less prominent if we initially start from the lower-frequency coupling in either constellation.

\section{Chains of coupled magnets}

\begin{figure}[htpb!]
    \centering
    \begin{subfigure}{0.5\columnwidth}
      \centering
      \includegraphics[scale=0.25]{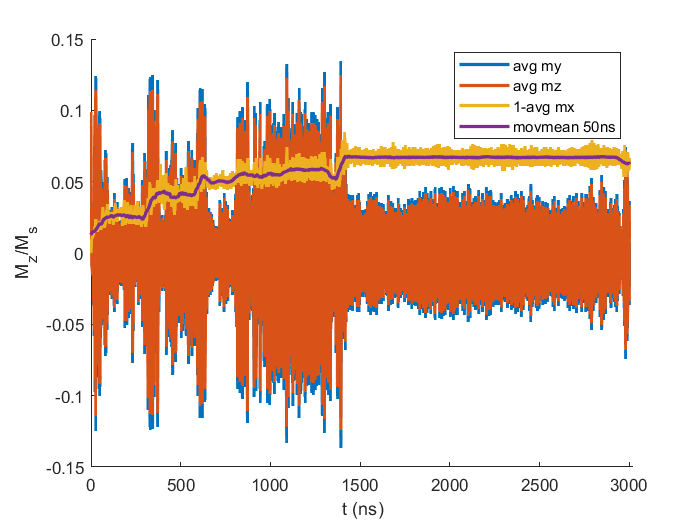}
      \caption{Average magnetization components of the whole chain.}
      \label{fig:chain_all_components}
    \end{subfigure}%
    \begin{subfigure}{0.5\columnwidth}
      \centering
      \includegraphics[scale=0.25]{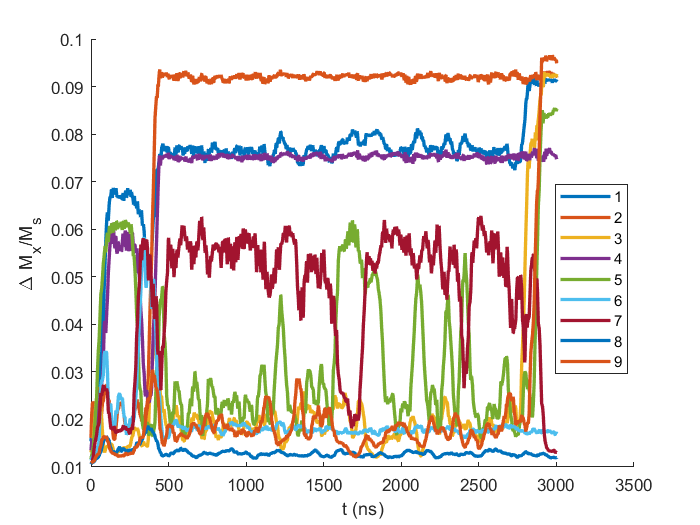}
      \caption{Oscillation amplitudes of each magnets of the chain. }
      \label{fig:chain_mx}
    \end{subfigure}
    \caption{Simulation of a chain of 9 magnets placed next to each other along their hard axis.}
    \label{fig:chain_excitation}
\end{figure}

Phase couplings between CNMPs can be observed for longer nanomagnet chains as well. 
Using a short chain of magnets result in quite more complex behaviour as the subfigures of Figure \ref{fig:chain_excitation} shows. First of all, the transient time is much longer, even at the end of this 3 $\mu$s simulation, there are prominent changes in some magnets' oscillations, whereas stand alone magnets or CNMPs had a couple hundred ns transient at maximum. During this transient the chain repeatedly stucks in local minima and then switches relatively fast to higher-amplitude oscillations overally creating a staircase-like increase of the average oscillation amplitude. As there are still changes at the end, it is not sure what the global optimum is.

By examining the pairwise phase of the neighbors, we concluded that these steps indicate either the synchronization of one or more pairs (possibly resulting in anti-synchronization of others) or coupling switches (in-phase to anti-phase or vice versa) which in overall can lead to a higher energy oscillation of the whole chain. 

We also visualized a few periods of the chain's oscillation in at least temporarily stable states. From these movies we concluded that in-phase couplings (what produce standing wave parts of the chain oscillation) usually occur at the ends of the chain with low-amplitudes and they are more often for lower excitation frequencies. Inversely, anti-phase couplings (what produce cuts in the chain oscillation) are more common in the middle with high amplitudes and they are more often for higher frequency excitations.

\section{Conclusions: toward computing devices}

In this paper we explored the rich dynamics of CNMPs - showing that in parametrically excited, coupled nanomagnets the phase of oscillations can be determined by field-coupling between the magnets. We envision that these results may be significant both from a fundamental physics and as application point of view.

On a physics level, CNMPs may be viewed as a phase-domain implementation of spin ices - where couplings (potentially competing couplings) give rise to complex magnetization patterns \cite{ref:adeyeye, ref:jack}. Spin ices are widely studied model systems of ordering and frustration,  arrays of CNMPs are expected to show similarly complex behaviors.  CNMPs likely offer more 'knobs' to control the couplings by tuning the parametric excitation frequency. We hope that this computational study inspires experimental investigation of the rich phase dynamics of nanomagnet arrays.

On the application side, CNMPs are one implementation of phase-based Ising systems \cite{ref:phase_ising}, which are intensely studied to solve computationally hard problems.

\begin{acknowledgments}

The authors acknowledge financial support from the Horizon 2020 Framework Program of the European Commission under FET-Open grant agreement no. 899646 (k-NET).

\end{acknowledgments}

\section*{Data Availability Statement}

The data that support the findings of this study are available from the corresponding author upon reasonable request.




\end{document}